\documentclass[11pt,tighten]{article}
\usepackage{epsfig,cite}

\begin{document}
\title{{\Large Emergence of highly--designable protein--backbone conformations in an off--lattice model }}
\author{Jonathan Miller, Chen Zeng$^*$, Ned S. Wingreen, and
Chao Tang$^\dag$ \\
NEC Research Institute, 4 Independence Way, Princeton, NJ 08540
USA}
\maketitle

\vspace{2cm}
\noindent
{\bf Classification:} {\it Biological Sciences}: Biophysics

\noindent
{\bf Corresponding author:} Chao Tang, NEC Research Institute, 4
Independence Way, Princeton, NJ 08540. Tel: (609)-951-2644; Fax:
(609)-951-2496.\\E-mail: tang@research.nj.nec.com


\pagebreak

\begin{center}
{\bf \Large Abstract}
\end{center}

Despite the variety of protein sizes, shapes, and backbone
configurations found in nature, the design of
novel protein folds remains an open problem. Within simple lattice models
it has been shown that all structures are not equally suitable for design.
Rather, certain structures are distinguished by unusually high
{\it designability}:
the number of amino--acid sequences for which they represent the unique
ground state; sequences associated with such structures possess both robustness to mutation
and thermodynamic stability. Here we report that highly designable
backbone conformations also emerge in a realistic off--lattice model.
The highly designable conformation of a chain of 23
amino acids are identified, and found to be remarkably insensitive to
model parameters. While some of these conformations correspond closely to 
known natural protein folds, such as the zinc finger and the helix-turn-helix
motifs, others do not resemble known folds and may be
candidates for novel fold design.

\pagebreak

{\bf \Large Introduction}

The de novo design of proteins--an object of enormous activity in recent years
\cite{Beasley97,Baltzer98,Cao98,Giver98,Regan98,Schafmeister98,Shakhnovich98,
Degrado99}--has so far dealt primarily with the {\it re}design of known protein
folds. Two major accomplishments in the direction of designing a fold
that is distinct from known natural folds are the
synthesis of a right-handed coiled coil \cite{Harbury98} and the synthesis of a
zinc finger without zinc \cite{Struthers96,Dahiyat97a,Dahiyat97b}. 
To challenge the best efforts of de novo design, nature offers roughly
1000 qualitatively distinct protein folds \cite{Chothia92}. 
Why has it proven difficult to design new protein folds? What program
should we follow to achieve ab-initio design of novel folds?

The principle of designability
\cite{Li96,Li98,Govindarajan95,Finkelstein87,Yue95} offers an answer to
both these
questions for simple lattice models. The designability of a structure is
measured by the number of sequences that design it, i.e. the
number of sequences that have the given structure as their unique lowest
energy conformation. Structures can differ vastly in their designability
\cite{Li96}, and it has been demonstrated that high designability entails
other protein-like properties, such as mutational stability, thermodynamic stability
\cite{Li96,Li98}, and fast folding kinetics \cite{Govindarajan95,Melin99}.
Design is {\it hard} in the sense that most structures have low
designability and their associated sequences lack these protein-like properties. For successful de
novo design, one should first identify the few highly designable structures.

It is an open question whether designability applies to real proteins as
it does to lattice polymers.
Real protein structures have a degree of complexity that
cannot be effectively represented within a simple lattice model.
For example, on a lattice the angles between bonds differ
from those naturally adopted in real proteins. Also, whereas
in a cubic lattice model the
cube minimizes surface area for a given volume and is
perfectly packed,
there exists no counterpart of the perfect cube once the 
lattice is removed.
For designability to guide practical design of new folds it
must apply to realistic descriptions of protein structure.

In this paper we report the computation of designability
within an off--lattice model that incorporates angles favored by natural proteins,
for protein chains of up to $N=23$ amino acids.
We find that the essential qualitative features of designability survive
the transition from lattice model to off--lattice model.
In particular, it remains true that a small fraction of 
compact structures are highly designable: these are nondegenerate
ground states
for an enormous number of amino--acid sequences. The
vast majority of structures, on the other hand, are suitable ground states
for few, if any, amino--acid sequences. Furthermore, the
sequences that fold into highly designable structures have enhanced thermodynamic stability --
the energy of the nearest excited state is separated from
the ground--state energy by an appreciable gap.

{\bf \Large Results}

{\bf Model}

Our off--lattice model is a $3$-state discrete--angle model of
the kind introduced by Park and Levitt \cite{Park95},
supplemented by uniform spheres centered on $C_{\alpha}$ and/or $C_{\beta}$
positions, in order to account for excluded volume effects. The energy of
a particular amino--acid sequence folded into a particular
backbone configuration is evaluated as the vector--product
of the hydrophobicity of the sequence dotted with the (normalized)
accessible surface area of each amino acid in the chain\cite{Flower97}.

{\bf Designability for a 23-mer}

The {\it designability} of a structure denotes the number
of distinct HP--sequences having that structure as
their unique ground state. Designability is an important attribute
of a structure, since it quantifies how many mutations
an amino--acid sequence can sustain while still folding
to the given ground--state configuration.

The distribution of designabilities for our model,
displayed in Fig.~\ref{histo},
reproduces a crucial feature first observed on the lattice:
While the vast majority of structures have very {\it low} designability,
the trailing edge (or tail) of the distribution 
consists of a small number of structures of very {\it high} designability.
Thus designability distinguishes a small subset of structures
from generic ones.

It turns out that
the identities of these highly designable structures
depend only weakly on the values of the
parameters that enter our calculation: the surface area cutoff
$A_c$, clustering radius $\lambda$, sidechain radius $r_\beta$,
and the set of allowed dihedral angles, and the range of 
amino--acid hydrophobicities.
More specifically, a significant fraction of structures
identified as highly designable for one set of parameter values
remains highly designable when these parameters are varied.
We provide evidence for this important observation in the
next five subsections.

{\bf Surface area cutoff}

As described in {\it Methods}, open structures are expected
to exhibit low designability. We anticipate that the highly
designable structures
of interest to us will fall mainly
within the class of compact structures,
and therefore only these compact structures are needed
in our calculation. The surface area cutoff $A_c$
determines how compact a structure must be in order
to qualify. We expect that, provided the choice of $A_c$
is not too restrictive, its particular value ought not
to be important.

A computationally practical choice of the surface--area cutoff eliminates
most of the less compact configurations. A few of these might
have proven highly designable if retained; however our objective
is not to find {\it all} highly designable structures, but only to identify
some of them. Therefore, our major concern is not that we might
incorrectly discard a few designable structures, but rather that we
might produce false positives:
structures that appear to be highly designable with a restrictive
value of the cutoff but have low designability for a
more relaxed cutoff. A larger cutoff admits previously
disallowed configurations that \lq\lq steal\rq\rq\ some sequences
from a configuration originally identified as highly
designable thereby reducing its designability.

In practice, as shown in Fig.~\ref{parameter}{\it a}, highly
designable structures tend to remain highly designable with
increasing surface--area cutoff. For example, 9 of the 10 most designable
structures remain within the 100 most designable even after
the surface--area cutoff is relaxed sufficiently to admit
a 10-fold increase in the number of participating structures.

{\bf Clustering radius}

As discussed in {\it Methods}, structures whose backbones differ
insignificantly from one another ought not to be considered distinct.
This observation is embodied in our calculation by grouping
into clusters those
structures whose backbone configurations lie within a certain crms
distance,  $\lambda$, of one another.
Varying the clustering radius, $\lambda$, leaves
unchanged the set of configurations that participate in the calculation.
For $\lambda \le 0.1 \AA$, nearly every cluster
consists of a unique configuration. To exhibit the dependence
of the most designable structures on $\lambda$, we fix a
configuration, and follow the designability of the cluster to which that
configuration belongs, as a function of $\lambda$. As shown in
Fig.~\ref{parameter}{\it b}, the most designable structures remain roughly the
same as $\lambda$ is varied over a wide range.

{\bf Sidechain radius}

Excluded--volume is incorporated by means of
a hard sphere of radius $r_\beta$ centered on the $\beta$-carbon
of each amino acid.
Increasing the sidechain radius $r_\beta$ eliminates some configurations
because of steric clashes, while decreasing $r_\beta$ 
admits previously ineligible configurations. Starting at $r_\beta=1.9
\AA$, we identify the most designable structures
and then count the fraction of these structures which remain highly
designable as $r_\beta$ is reduced. As shown in Fig.~\ref{parameter}{\it c},
the identities of the most designable structures are well--preserved.

{\bf Choice of angles}

Next, we address to what extent an outcome depends on
a particular choice of the discrete set of dihedral angles.
A discrete set of
angles cannot sample the structure space fully, and so cannot
``hit'' all possible structures. On the other hand, we know that
the designability of a
structure depends on the local density of solvent--exposure
vectors $\tilde{\bf A}$\cite{Li98}--with highly
designable structures occupying the lowest density regions. If the subset of
structures sampled by a discrete set of angles reasonably preserves
{\it density} in the space of structures, 
highly designable structures should remain highly
designable as we improve our sampling of structure space.

To examine this possibility,
we identify configurations generated by one angle set and follow
their cluster designabilities as configurations from other
angle sets are added. We take five different angle sets derived
from fitting to 1PSV, and use the most compact configurations generated
by each set.
We calculate the designability of structures using configurations
from, respectively, one, two, three, four, and finally all
five sets. We observe in Fig.~\ref{parameter}{\it d} that the most designable
structures in set \#1 remain highly designable even as configurations
from sets \#2, \#3, \#4, and \#5 are added. This result is maintained
under permutation of the five sets. Apparently,
any reasonable choice of angle set covers the structure
space sufficiently well that highly designable structures can be
identified with high probability.

{\bf HP sequences }

To check whether the identification of designable structures
depends on our use of HP (binary) sequences of amino acids,
we recalculate
designabilities using amino acids with continuous real-valued
hydrophobicities.
We randomly choose 4,000,000 sequences ${\bf h}=(h_1,\cdots,h_N)$, where
$h_i \in [0,1]$, and evaluate their energy for all configurations using
equation~(\ref{ham}). In Fig.~\ref{parameter}{\it e} we plot the designability
calculated this way against that from the enumeration of HP sequences.
As the figure shows, the highly designable structures computed
by these two alternative methods are nearly identical.

{\bf Parameter Independence}

In the preceeding five subsections we have demonstrated
that the parameters can sustain a considerable degree of variation
without significantly changing the outcome of the
designability calculation. The weak dependence
of the set of highly designable structures on parameters
is illustrated in Fig.~\ref{parameter}. Because the
identity of the highly designable structures
is robust to parameter variation, we now 
examine their potential as candidates for design.

{\bf Gap}

In particular, a prerequisite for design is believed to be the
presence of a large separation between the ground--state energy and the
energy of the lowest excited state. For each structure, we have identified
the HP-sequence that makes this gap the largest. The value of this largest
gap is shown in Fig.~\ref{gap}, as a function of the designability of
the structure.

To convert the vertical scale of Fig.~\ref{gap} to
real energies,
we observe that one unit of energy corresponds to a sequence
of exclusively hydrophobic amino acids $(h_i=1)$ folded
into one of our typical compact structures.
Our choice of surface area cutoff $A_c$
guarantees that a typical compact configuration 
has around half of its maximal
accessible surface exposed - about $25 \AA^2$ 
per residue.
A conservative estimate for the energy of exposed surface,
$20$~cal$/\AA^2/$mol\cite{Creighton} then yields
an energy on the order of $10$~kcal$/$mol for a $23$-mer.
The highest gap energies achieved in Fig.~\ref{gap}, of order
$0.05$, therefore correspond to a gap of $0.5$~kcal$/$mol, around $k_B T$
for room temperature. This gap is roughly the energy to
promote one hydrophobic amino--acid  from core to surface,

Also plotted is the average gap for all HP-sequences which
design a structure. It is evident that high designability correlates
strongly with a large gap.

{\bf \Large Discussion}

{\bf Designability off--lattice}

The principle of designability is that some protein structures
are intrinsically easier to design than others. However,
up to now, designability has been demonstrated only in highly
restrictive lattice models. 
Our calculations indicate that the qualitative features of
designability in lattice models are also exhibited off--lattice.
Namely, a small minority of off-lattice structures are
distinguished by high designability: these structures
are lowest--energy states for many more than their share
of sequences. Moreover, the sequences associated with
these structures have enhanced thermodynamic stability.
The work presented here, using a realistic off--lattice model for
protein--backbone configurations, makes it more plausible that designability
applies to real proteins.

{\bf Highly designable structures}

The insensitivity to model parameters of the results presented
suggests that our highly designable
structures are possible candidates for real protein design.
 It is therefore worthwhile
to study some of our best candidates in detail, and to understand what 
architectural properties distinguish the most designable structures
from the least designable ones, and how the most designable ones compare
with known natural structures.

\begin{figure}
\centering
\epsfig{file=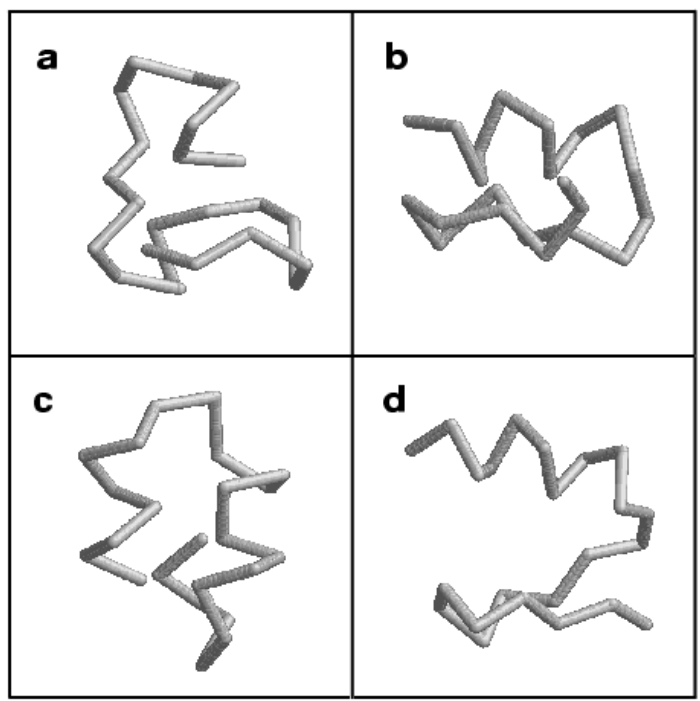,width=9.3cm}
\vspace{5cm}
\caption{Miller, et al.}
\label{configs}
\end{figure}

\begin{figure}
\centering
\epsfig{file=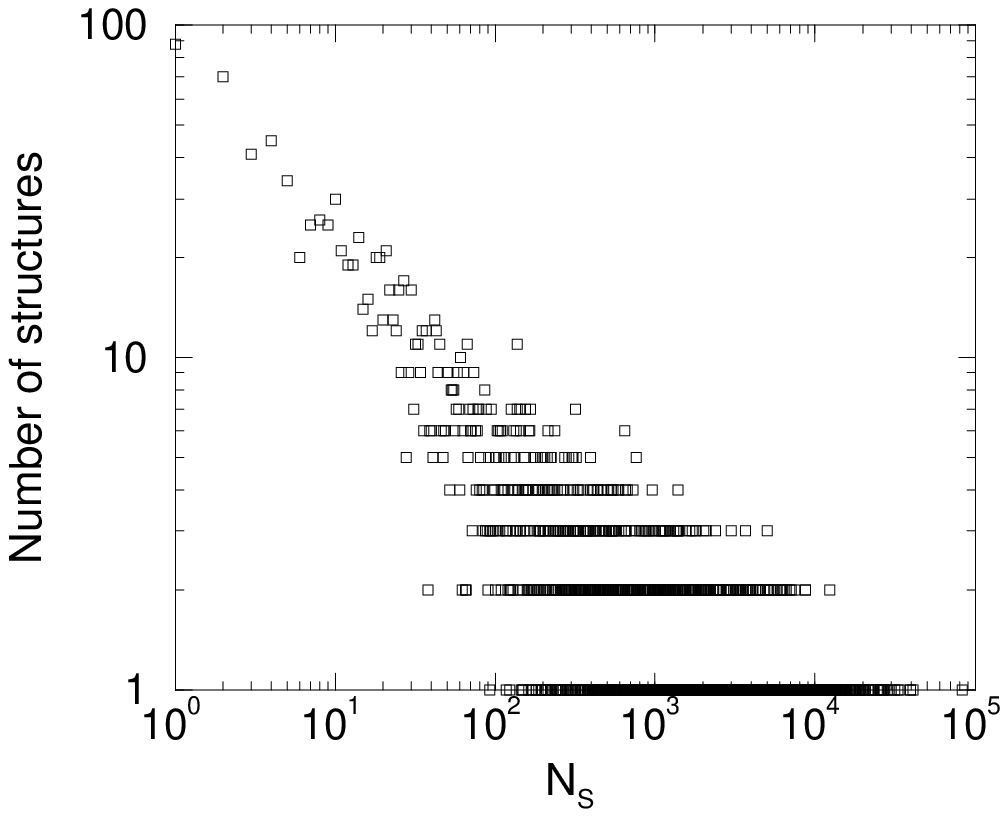,width=8.7cm}
\vspace{5cm}
\caption{Miller, et al.}
\label{histo}
\end{figure}

\begin{figure}
\centering
\epsfig{file=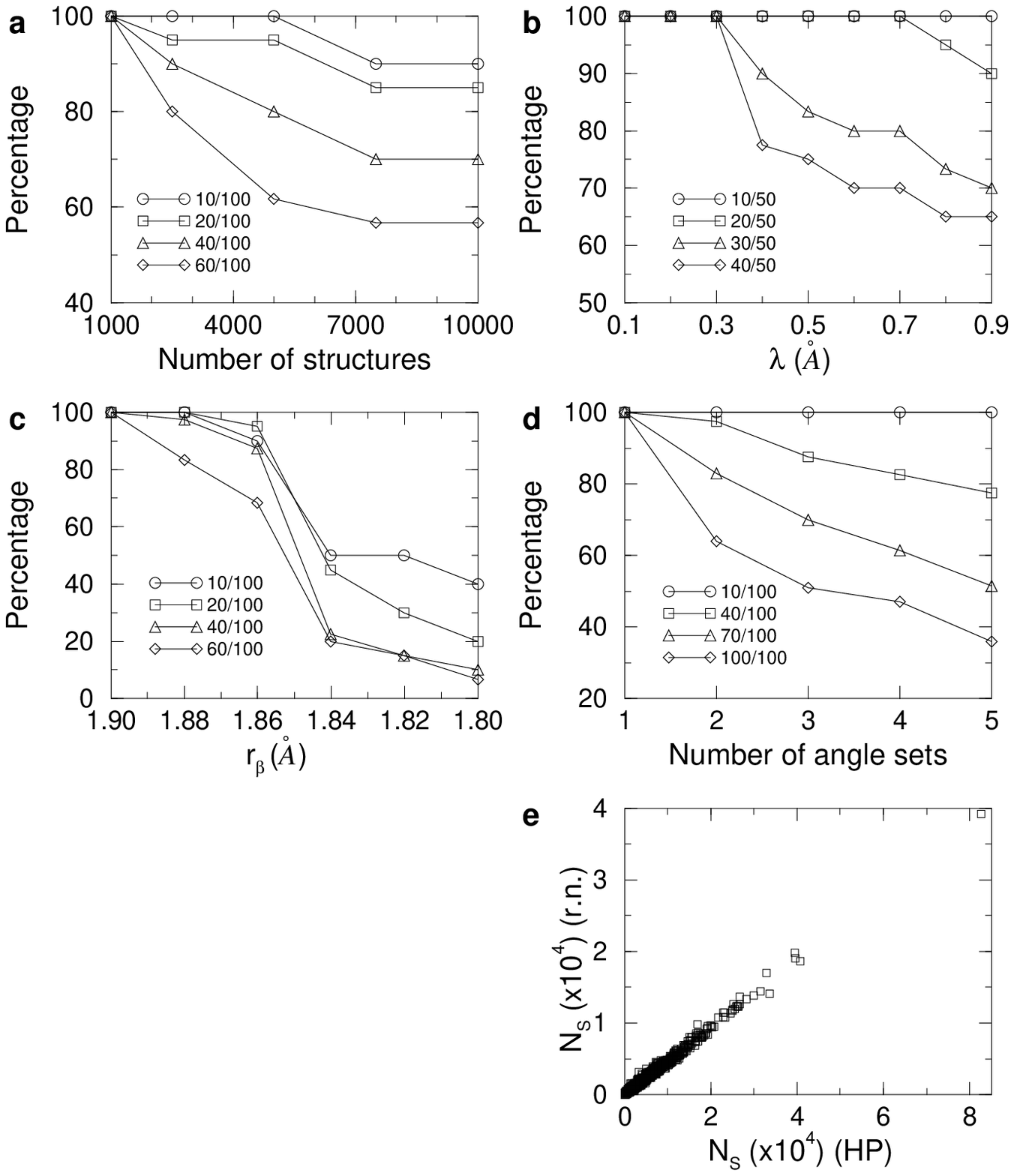,width=12cm}
\vspace{3cm}
\caption{Miller, et al.}
\label{parameter}
\end{figure}

\begin{figure}
\centering
\epsfig{file=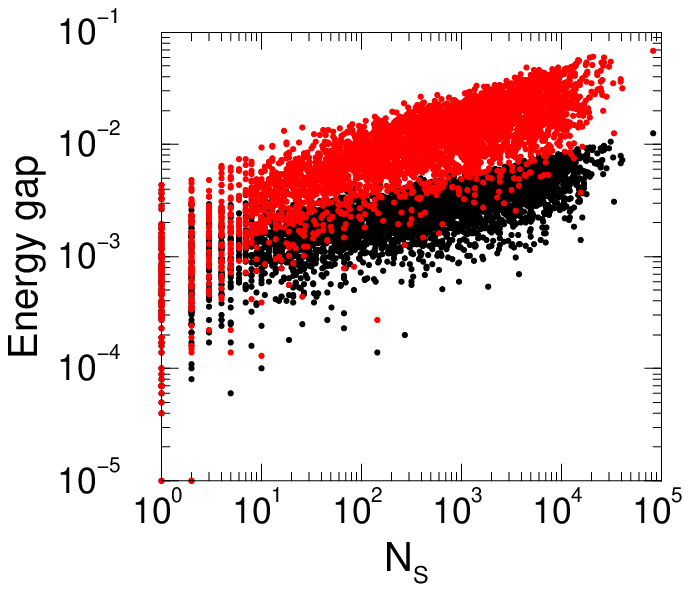,width=8.7cm}
\vspace{5cm}
\caption{Miller, et al.}
\label{gap}
\end{figure}

\begin{figure}
\centering
\epsfig{file=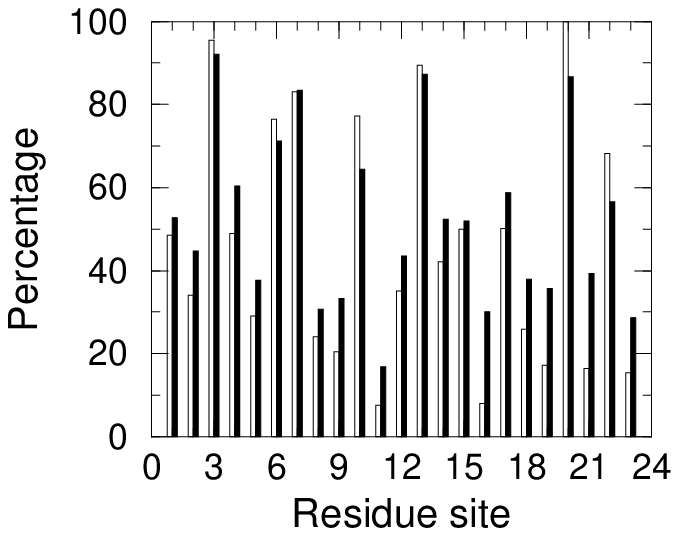,width=8.7cm}
\vspace{5cm}
\caption{Miller, et al.}
\label{expos}
\end{figure}

\begin{figure}
\centering
\epsfig{file=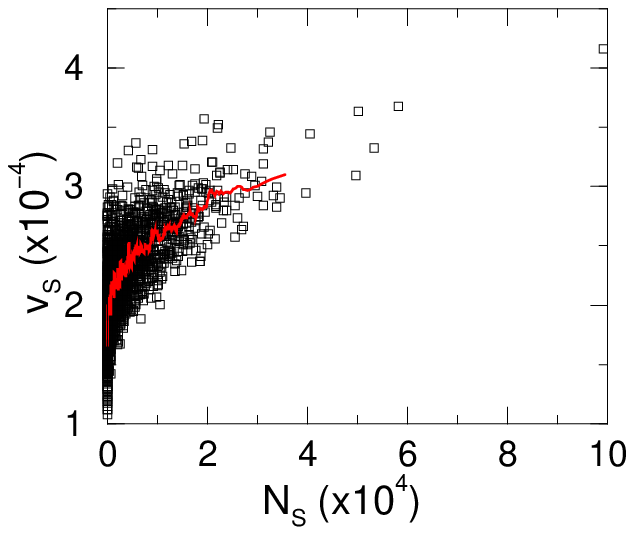,width=8.7cm}
\vspace{5cm}
\caption{Miller, et al.}
\label{vari}
\end{figure}

Representative configurations of some of the most designable structures
are shown in Fig.~\ref{configs}{\it a}-{\it c}. A striking characteristic of
the highly designable structures is that each has a well-defined core
consisting of a small subset of the amino acids of the chain. For example, in
Fig.~\ref{expos} we have plotted the inaccessible surface area of each
amino acid along the chain for the configuration appearing in
Fig.~\ref{configs}{\it b}. Observe that 5 of the 23 amino acids are more than
70\% buried. Also shown in Fig.~\ref{expos} is the probability that a
hydrophobic amino acid occupies a particular site, averaged over
all HP--sequences that
design the structure, revealing the preference of hydrophobic amino acids 
for the core.
A quantitative measure of the core in a structure is
the variance $v_S$ of the exposure vector $\tilde{\bf A}$: $v_S = (1/N)
\sum_i \tilde{a}_i^2 - (1/N^2) (\sum_i \tilde{a}_i)^2$. In Fig.~\ref{vari},
we plot $v_S$ versus the designability $N_S$.  On average
the two quantities correlate well; however, the scatter of the data is
large in the region of low $N_S$: structures with well-formed cores are
not necessarily highly designable. 

A zinc-finger-like fold emerges from our calculation as one
of the most designable structures. The fold (Fig.~\ref{configs}{\it b})
does not simply replicate 1PSV (Fig.~\ref{configs}{\it d}), on which we
optimized our angle set. The structure of 1PSV is too open to be
designable within our model because the small, uniformly--sized
sidechains cannot fill the large opening between the 
$\alpha$-helix and the $\beta$-$\beta$ turn in 1PSV.
Interestingly, the model produces a highly designable solution by collapsing the
$\alpha$-helix onto the $\beta$-$\beta$ turn.

Another of our most designable structures is similar to
another small natural fold, the helix-turn-helix
(see Fig.~\ref{configs}{\it c}). 

Some of our
most designable structures (e.g., that
shown in Fig.~\ref{configs}{\it a}) do not resemble any
known natural folds. These structures are candidates for the design of truly
novel folds. 

\begin{figure}
\centering
\epsfig{file=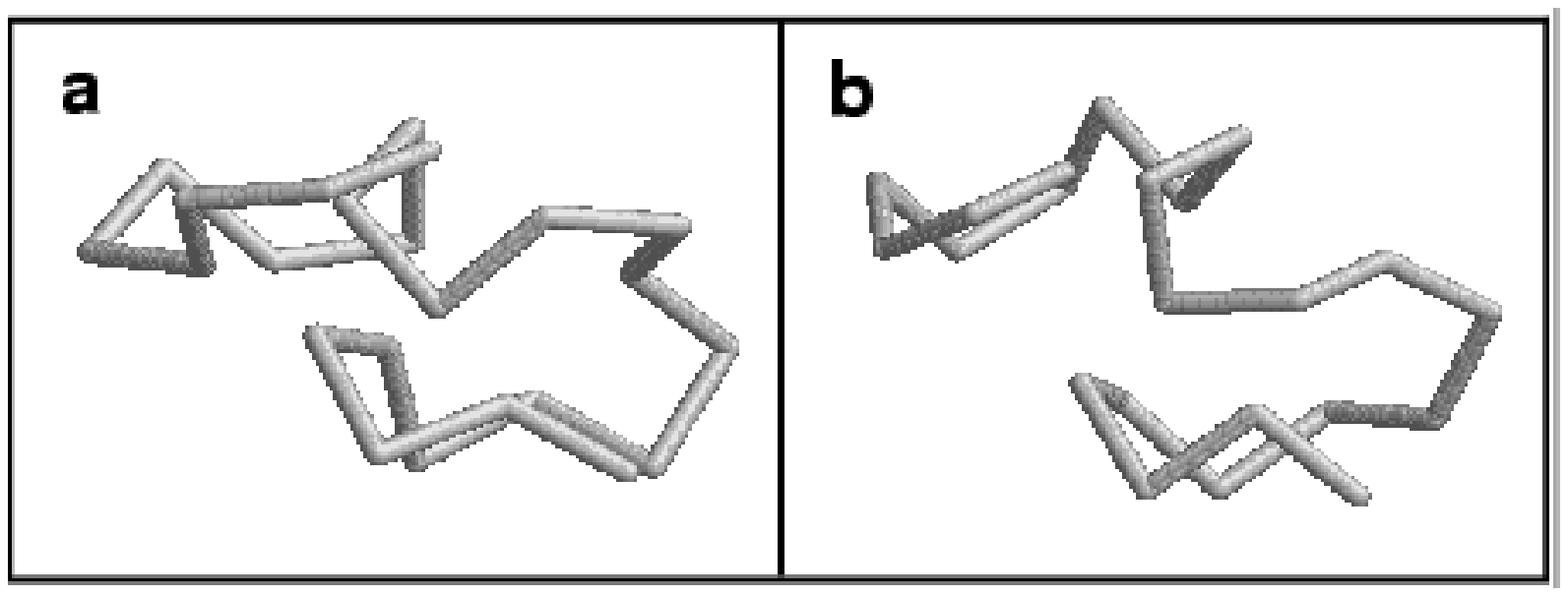,width=8.7cm}
\vspace{5cm}
\caption{Miller, et al.}
\label{zfls}
\end{figure}

Targeting a fold by fitting the angle set to a chosen structure is not
essential. For example, we can obtain a suitable angle set by choosing
two pairs of dihedral angles $(\phi,\psi)$ within the $\beta$-sheet
region and one pair from the $\alpha$-helix region, locally
optimizing on $160$ representative natural structures from the PDB
database \cite{Park95}.
Among the most designable structures emerging
for this angle set is the zinc-finger-like structure
in Fig.~\ref{zfls}{\it a}, shown next to its apparent natural counterpart,
1NC8 \cite{Kodera98} (Fig.~\ref{zfls}{\it b}).

{\bf \Large Conclusions}

In summary, we have computed the designabilities of structures within an
off-lattice model of realistic protein--backbone configurations. 
Highly designable structures emerge with remarkable insensitivity
to model parameters. The sequences which design these structures have
strongly enhanced
mutational stability and a large energy gap between the native fold
and the lowest non-native conformation.
In this light, it is interesting that recent mutation studies on 
some small proteins show that they maintain their native folds
even when about half of their residues
are replaced by alanine \cite{Kuroda00,Brown99}. Some of our highly designable
structures correspond closely to natural folds, such as the zinc--finger
and helix--turn--helix motifs. Others do not resemble existing structures,
and are candidates for ab-initio design of novel protein folds.

{\bf \Large Methods}

{\bf Model}

The model we adopt is closely related to the off-lattice,
$m$-state discrete-angle model introduced by Park and Levitt \cite{Park95}.
Each configuration is defined by a sequence of $C_{\alpha}$ bonds of
length $3.8\AA$, and each pair of dihedral angles $(\phi,\psi)$
is restricted to one of only $m$ alternatives; here we take $m=3$.
The set of $m$ allowed angle pairs is chosen
by fitting to the backbone
coordinates of representative natural proteins\cite{Park95}, as discussed
below. To suppress
self-intersections of the chain, we augment the model by introducing a
volume for the amino-acid residues in the form of a sphere
of radius $r_{\beta}$ centered on $C_{\beta}$ (the first carbon of the
sidechain). The backbones of some configurations constructed in this
fashion are shown in Fig.~\ref{configs}{\it a}-{\it c}.

This off-lattice model incorporates properties of real polymers not well
reproduced in simple lattice models. On the lattice, for example, allowed
ground-state structures were limited to those maximally compact structures
that fill the unique rectangle or box of minimum surface area. Off the
lattice, every structure can be expected to have a distinct surface area,
but once again, open or extended structures are not expected
to be designable.
We entertain as plausible ground-state structures only those with a surface
area below some cutoff value $A_c$, which enters our computation as a
parameter$^\ddag$. 

Because a discrete angle set represents only a crude approximation to a continuum
of angles, it is unrealistic to expect the surface area of a
discrete-angle structure to faithfully reproduce the surface area of a structure
built from more flexible angles.
Importantly, using flexible angles
would allow our more open structures, {\it e.g.} those just
below the cutoff $A_c$, to contract and reduce their
exposed surface areas. To achieve this equalizing effect
of a continuum of angles within the limitations of
a discrete--angle model,
we normalize the vector of solvent-accessible
surface areas ${\bf A} = (a_1,\cdots, a_N)$, where $a_i$ is the
solvent-accessible surface area of the $i$-th residue, in such a way
as to preserve the {\it pattern} of surface exposure along a chain.
A suitable procedure$^\S$ is to normalize the vector ${\bf A}$ for
each structure by the total exposed surface area of that structure:
${\tilde{\bf A}} = {\bf A}/\sum_i a_i = (\tilde{a}_1,\cdots,\tilde{a}_N)$.
This procedure treats all structures 
below the cutoff $A_c$ as equally compact, while preserving
each structure's individual pattern of surface exposure
along the chain.

{\bf Clustering}

As with real proteins, description and comparison of configurations
off-lattice demands precision about what we mean by the term ``structure.''
For example, a protein structure obtained by NMR represents an ensemble
of configurations, no element of which necessarily provides a better
fit to the data than any other. This ensemble presumably reproduces
the temperature--induced fluctuations of a natural protein around its 
native state.
On averaging over this ensemble for small stably-folded polypeptides in the PDB database,
one finds a typical
crms of roughly $0.3-0.5 \AA$ per residue. A similar range of crms can
be inferred from the $B$ values of protein crystals\cite{Creighton}.
Accordingly, our off-lattice polymer configurations are grouped into 
clusters consisting of all configurations lying within 
a crms distance $\lambda$ per residue of one another.
Configurations within a cluster are to be thought of as variations of
a single structure, and we refer to clusters and
structures interchangeably. 

{\bf Designability}

We define the designability of a structure as the sum of the
designabilities of its included configurations. The designability of a
configuration is simply the number of sequences with that configuration
as a unique ground state\cite{Li96,Li98}. To evaluate the energy of a
sequence on each configuration, we associate a hydrophobicity $h_i$ with
each amino acid of the sequence. In practice, we assign a hydrophobicity
which is either 0 (Polar) or 1 (Hydrophobic) to each monomer to
create an HP--sequence\cite{HP}; that this is a reasonable
simplification finds support in the work of Hecht and
co--workers\cite{Beasley97}
({\it cf.} Fig.~\ref{parameter}{\it e} for the results of a more general choice).
The energy of a particular sequence folded into a particular configuration
is obtained by taking the sum of the products of each amino acid's
hydrophobicity $h_i$  with its normalized surface exposure $\tilde{a}_i$,
\begin{equation}
E = \sum_i h_i \tilde{a}_i.
\label{ham}
\end{equation} 
We numerically evaluate the energy of all HP--sequences for all configurations.

{\bf Parameters}

Except as indicated explicitly in the text, we have chosen discrete angles
and the amino--acid radius to optimize the fit to the backbone of 
the zinc--less synthetic zinc finger 1PSV\cite{Dahiyat97b}
(Fig.~\ref{configs}{\it d}). We find
that there are many angle sets that fit the backbone of 1PSV
almost equally well. For example, the crms per residue between 1PSV and
the structure obtained from each of our 10 best angle sets
varies from $0.844 \AA$ to $0.913 \AA$. The angle
set we use for most of the calculations presented in this paper is
$(\phi,\psi)=(-95^\circ,135^\circ)$ ($\beta$--region),
$(-75^\circ,-25^\circ)$ ($\alpha$--region), and
$(-55^\circ,-55^\circ)$ ($\alpha$--region). We take $r_{\beta}=1.9\AA$, the
radius above which the amino acids fit to the backbone of 1PSV
would clash.

\newpage

\begin{center}
{\bf Footnotes}
\end{center}

\noindent
* Present address: Department of Physics, George Washington University,
Washington, D.C. 20052, USA.

\noindent
\dag \ To whom reprint requests should be addressed. E-mail:
tang@research.nj.nec.com.

\noindent
\ddag \ We evaluate the area of each $C_{\beta}$ sphere accessible to a probe
sphere of radius $1.4 \AA$, by the methods used in the program SERF\cite{Flower97};
the slightly different values of surface area
obtained by the different methods do not in any way alter the
outcome of the calculations.

\noindent
\S \ We have checked that certain alternative normalizations (for example,
normalizing by the total solvent-{\it inaccessible} surface area) do not
alter the set of highly designable structures that emerge from our
calculation. With {\it no} normalization, higher designability becomes closely
correlated with lower solvent-accessible surface area.

\newpage

\noindent{\bf Figure Captions}
\begin{itemize}

\item[1.] ({\it a})-({\it c}) Backbone configurations of 1{\it st},
4{\it th}, and 15{\it th} most designable 23-mer structures. ({\it d})
Backbone configuration of the zinc finger 1PSV\cite{Dahiyat97b},
truncated to 23 amino acids.

\item[2.] Histogram of designabilities of 23-mer structures, using
$r_\beta = 1.9 \AA$. The surface area cutoff $A_c$ is such that 10,000
configurations participate in the calculation,  grouped into
4688 clusters with cluster radius $\lambda=0.4 \AA$. 

\item[3.] Sensitivity to parameter changes of the most designable
structures from Fig.~\ref{histo}.
({\it a}) Fraction of the 10, 20, 40, or 60
most designable structures which remain in the 100 most designable
as the surface--area cutoff increases. The initial cutoff $A_c$ is chosen
so that only the 1000 most compact configurations participate
and $A_c$ increases until 10,000 configurations participate.
({\it b}) Fraction of the 10, 20, 30, or 40 most designable structures
which remain in the  50 most designable as the clustering radius
$\lambda$ is increased.
The 5000 most compact configurations participate in the calculation and 
$r_\beta = 1.9 \AA$.
({\it c}) Fraction of the 10, 20, 40, or 60 most designable
structures which remain in the 100 most designable as the
sidechain radius $r_\beta$ is changed. We have
chosen the surface  area
cutoff so that 5000 structures participate in the designability 
calculation for $r_\beta=1.9\AA$.  If some configurations of the
original most designable structures are not among the 5000 most
compact configurations
for some smaller $r_\beta$, we nevertheless retain them in the calculation.
The clustering radius is
$\lambda=0.4 \AA$.
({\it d}) Fraction of the  10, 40, 70, or 100 most designable
structures which remain in the 100 most designable as configurations
from other angle sets are added. The values of the five angle
sets are: set $\#1 =
(-95^\circ,135^\circ)$, $(-75^\circ,-25^\circ)$, $(-55^\circ,-55^\circ)$;
set $\#2 = (-95^\circ,135^\circ)$, $(-85^\circ,-55^\circ)$,
$(-65^\circ,-25^\circ)$; set $\#3 = (-105^\circ,145^\circ)$,
$(-85^\circ,-15^\circ)$, $(-75^\circ,-35^\circ)$; set $\#4 =
(-105^\circ,145^\circ)$, $(-85^\circ,-35^\circ)$, $(-85^\circ,-5^\circ)$;
set $\#5 = (-105^\circ,145^\circ)$, $(-85^\circ,-35^\circ)$,
$(-85^\circ,-15^\circ)$.
({\it e}) Designability of structures obtained from 4,000,000
randomly generated sequences of real numbers in [0,1] versus designability
from
enumeration of HP--sequences. The 10,000 most compact configurations
participate in the calculation, $\lambda=0.4 \AA$, and $r_\beta = 1.9
\AA$.

\item[4.] Maximum energy gap (red dots) and average energy
gap (black dots) for the HP--sequences which design a given structure,
plotted versus structure designability. The 10,000 most compact
configurations of the 23-mer participate in the calculation, with
$\lambda=0.4 \AA$ and $r_\beta = 1.9 \AA$.

\item[5.] Solid bars: Inaccessible surface for residues ($C_{\beta}$
spheres) of the highly designable configuration shown in Fig.~1{\it b}. 
Hollow bars: Probability, averaged over all HP--sequences that design
the configuration, that ra particular site along the chain is occupied by a
hydrophobic amino acid.

\item[6.] The average variance $v_S$ of a cluster against the
designability $N_S$ of the cluster for the 23-mer. The 5000 most compact
configurations participate in the calculation, $\lambda=0.4 \AA$, and
$r_\beta = 1.9 \AA$. Red line: running average with bin size 30.

\item[7.] ({\it a}) Backbone configuration of the 11{\it th}
most designable 23-mer structure, using untargeted angle
set (see text):
$(\phi,\psi)=(-55^\circ,135^\circ)$, $(-126^\circ,145^\circ)$, and
$(-85^\circ,-25^\circ)$, with a mean crms of $3.6\AA$ on a representative
subset of natural structures segmented into subchains of 21 amino acids.
For this calculation, the amino acids are
represented by spheres of radius $r_{\alpha}=1.52\AA$ centered
on the $C_{\alpha}$ carbons only. ({\it b})
Backbone configuration of the zinc finger 1NC8\cite{Kodera98},
truncated to 23 amino acids.

\end{itemize}

\newpage

\newpage

\newpage

\newpage

\newpage

\newpage

\newpage

\newpage

\begin{thebibliography}{99}

\bibitem{Beasley97}
Beasley,~J.~R. \& Hecht,~M.~H. (1997). Protein design: The choice of de novo sequences.
{\it J. Biol. Chem.} {\bf 272}, 2031-2034.

\bibitem{Baltzer98}
Baltzer,~L. (1998). Functionalization of designed folded polypeptides.
{\it Curr. Opin. Struct. Biol.} {\bf 8}, 466-470.

\bibitem{Cao98}
Cao,~A.~N., Lai,~L.~H. \& Tang,~Y.~Q. (1998).
The current state and prospect of de novo protein design.
{\it Prog. Biochem. Biophys.} {\bf 25}, 197-201.

\bibitem{Giver98}
Giver,~L. \& Arnold,~F.~H. (1998). Combinatorial protein design by in vitro recombination.
{\it Curr. Opin. Chem. Biol.} {\bf 2}, 335-338.

\bibitem{Regan98}
Regan,~L. \& Wells,~J. (1998). Engineering and design: recent adventures in molecular design - Editorial overview. {\it Curr. Opin. Struct. Biol.} {\bf 8}, 441-442.

\bibitem{Schafmeister98}
Schafmeister,~C.~E. \& Stroud,~R.~M. (1998). Helical protein design.
{\it Curr. Opin. Biotechnol.} {\bf 9}, 350-353.

\bibitem{Shakhnovich98}
Shakhnovich,~E.~I. (1998). Protein design: a perspective from simple tractable models.
{\it Fold. Design}, {\bf 3}, R45-R58.

\bibitem{Degrado99}
DeGrado,~W.~F., Summa,~C.~M., Pavone,~V., Nastri,~F. \& Lombardi,~A. (1999).
De novo designa and structural characterization of proteins and metalloproteins.
{\it Annu. Rev. Biochem.} {\bf 68}, 779-819.

\bibitem{Harbury98}
Harbury,~P.~B., Plecs,~J.~J., Tidor,~B., Alber,~T. \& Kim,~P.~S. (1998).
High--resolution protein design with backbone freedom.
{\it Science}, {\bf 282}, 1462-1467. 

\bibitem{Struthers96}
Struthers,~M.~D., Cheng,~R.~P. \& Imperiali,~B. (1996).
Design of a monomeric $23$--residue polypeptide with defined tertiary structure.
{\it Science}, {\bf 271}, 342-345. 

\bibitem{Dahiyat97a}
Dahiyat,~B.~I. \& Mayo,~S.~L. (1997).
De novo protein design: fully automated sequence selection.
{\it Science}, {\bf 278}, 82-87.

\bibitem{Dahiyat97b}
Dahiyat,~B.~I., Sarisky,~C.~A. \& Mayo,~S.~L. (1997).
De novo protein design: towards fully automated sequence selection. {\it J. Mol. Biol.} {\bf
273}, 789-796.

\bibitem{Chothia92}
Chothia,~C. (1992).
Proteins - 1000 families for the molecular biologist.
{\it Nature}, {\bf 357}, 543-544.

\bibitem{Li96}
Li,~H., Helling,~R., Tang,~C. \& Wingreen,~N. (1996).
Emergence of preferred structures in a simple model of protein--folding.
{\it Science}, {\bf 273}, 666-669.

\bibitem{Li98}
Li,~H.,~Tang, C.~\& Wingreen, N.~S. (1998).
Are protein folds atypical?
{\it Proc. Natl. Acad. Sci. USA} {\bf 95}, 4987-4990.

\bibitem{Govindarajan95}
Govindarajan,~S. \& Goldstein,~R.~A. (1995).
Searching for foldable protein structures using optimized energy functions.
{\it Biopolymers}, {\bf 36},
43-51.

\bibitem{Finkelstein87}
Finkelstein,~A.~V. \& Ptitsyn,~O.~B. (1987).
Why do globular proteins fit the limited set of folding patterns?
{\it Prog. Biophys. Mol. Biol.} {\bf 50}, 171-190.

\bibitem{Yue95}
Yue,~K. \& Dill,~K.~A. (1995).
Forces of tertiary structural organization in globular proteins.
{\it Proc. Natl. Acad. Sci. USA} {\bf 92}, 146-150.

\bibitem{Melin99}
M\'elin,~R., Li,~H., Wingreen,~N.~S. \& Tang,~C. (1999).
Designability, thermodynamic stability, and dynamics in protein folding: a lattice
model study.
{\it J. Chem. Phys.} {\bf 110}, 1252-1262.

\bibitem{Park95}
Park,~B.~H. \& Levitt,~M. (1995).
The complexity and accuracy of discrete state models of protein structure.
{\it J. Mol. Biol.} {\bf 249}, 493-507.

\bibitem{Flower97} 
Flower,~D.~R. (1997).
SERF: A program for accessible surface area calculations.
{\it J. Mol. Graph. Model.} {\bf 15}, 238-244.

\bibitem{Creighton}
Creighton,~T.~E. (1993).
{\it Proteins} (Freeman, New York, ed. 2), pp. 236-237; pp. 160-162.

\bibitem{Kodera98} Kodera,~Y., Sato,~K., Tsukahara,~T., Komatsu,~H.,
Maeda,~T. \& Kohno,~T. (1998).
High--resolution solution NMR structure of the minimal active domain of the
human immunodeficiency virus type-2 nucleocapsid protein.
{\it Biochemistry}, {\bf 37}, 17704-17713.

\bibitem{Kuroda00}
Kuroda,~Y. \& Kim,~P.~S. (2000).
Folding of bovine pancreatic trypsin inhibitor (BPTI) variants in which
almost half the residues are alanine.
{\it J. Mol. Biol.} {\bf 298}, 493-501.

\bibitem{Brown99}
Brown,~B.~M. \& Sauer,~R.~T. (1999).
Tolerance of Arc repressor to multiple--alanine substitutions.
{\it Proc. Natl. Acad. Sci. USA} {\bf 96}, 1983-1988.

\bibitem{HP}
Lau,~K.~F. \& Dill,~K.~A. (1989).
Lattice statistical mechanics model of the conformational and sequence spaces
of proteins.
{\it Macromolecules}, {\bf 22}, 3986-3997. 

\end{thebibliography}
\end{document}